\documentclass[preprintnumbers,prd, amsmath,amssymb,a4page, latexsym,array,enumerate,letter,numbers, twocolumn]{revtex4}
\usepackage{graphicx}% Include figure files
\usepackage{dcolumn}% Align table columns on decimal point
\usepackage{bm}% bold math
\usepackage{amssymb}
\usepackage{epsfig}
\usepackage{slashed}
\usepackage{color}

\usepackage[
       colorlinks=true,
      filecolor=black,
       anchorcolor=blue,
      linkcolor=blue,
      citecolor=red,
      urlcolor=cyan,
       linktocpage=true,
        plainpages=false,
        breaklinks=true,
            pdfstartview=FitH
           ]{hyperref}
           
\newcommand{\nn}{\nonumber}

\def\tR{\tilde{R}}
\def\tF{\tilde{F}}

\def\va{{{a}}}%\varphi}}
\def\vt{{\vartheta}}
\def\ra{d_\va}
\def\rd{L}
\def\hg{{{g}}}
\def\ka{k_*}

\def\lf{\alpha_\gamma}
\def\lr{\alpha_{\hg}}

\def\L{\mathcal{L}}

\def\e{{e}}
\def\ci{\theta_i}
\def\eL{\varepsilon_L}
\def\eR{\varepsilon_R}
\def\ve{\varepsilon}

\begin{document}

%\setcounter{tocdepth}{2}
%\tableofcontents
%\title{Gravitational Waves and Possible Fast Radio Bursts from Axion Clumps}

\title{Fast Gravitational Wave Bursts from Axion Clumps}
%\title{Possible fast radio and gravitational wave bursts from axion clumps}

\author{Sichun Sun$^{1,2}$} 
\email[corresponding author: ]{sichunssun@bit.edu.cn}
%\email{Sun.Sichun@roma1.infn.it}
\author{Yun-Long Zhang$^{3,4,5}$}
\email[corresponding author: ]{zhangyunlong@nao.cas.cn}
%\email{yun-long.zhang@yukawa.kyoto-u.ac.jp }
\affiliation{$^1$School of Physics, Beijing Institute of Technology, Haidian District, Beijing 100081, China}

\affiliation{$^2$Department of Physics and INFN, Sapienza University of Rome, Rome I-00185, Italy}

\affiliation{$^3$National Astronomy Observatories, Chinese Academy of Science, Beijing, 100101, China}

\affiliation{$^4$School of Fundamental Physics and Mathematical Sciences, Hangzhou Institute for Advanced Study,
University of Chinese Academy of Sciences, Hangzhou 310024, China}

\affiliation{$^5$Center for Gravitational Physics, Yukawa Institute for Theoretical Physics,  
 Kyoto University, Sakyo-ku, Kyoto 606-8502, Japan   }

%\preprint{YITP-20-28}
\begin{abstract}
The axion objects such as axion mini-clusters and axion clouds around spinning black holes induce parametric resonances of electromagnetic waves through the axion-photon interaction. In particular, it has been known that the resonances from the axion with the mass around $10^{-6}$eV may explain the observed fast radio bursts (FRBs). Here we argue that similar bursts of high frequency gravitational waves, which we call fast gravitational wave bursts (FGBs), are generated from axion clumps with the presence of gravitational Chern-Simons (CS) coupling. The typical frequency is half of the axion mass, which in general can range from kHz to GHz. We also discuss the secondary gravitational wave production associated with FRB, as well as the possible host objects of the axion clouds, such as primordial black holes with typical masses around $10^{-5}M_{\odot}$. Future detections of FGBs together with the observed FRBs are expected to provide more evidence for the axion.
\end{abstract}
\maketitle

%\tableofcontents 
% the wavelength $10^ {26}$m (
\section{Introduction}

%{\it \bf Introduction. ---}
The first observation of the gravitational waves has started a new era for astrophysics, cosmology and even particle physics. Searching for gravitational wave signals in the universe at different frequencies from different sources has started, from the primordial gravitational waves at $10^{-16}$ Hz to the pulsars and binary system signals up to $10^4$ Hz. What's more, the higher frequency up to MHz or GHz gravitational wave searches have also been proposed \cite{Arvanitaki:2012cn,Ito:2019wcb}, and those signals can be generated by the axion annihilation around the black holes \cite{Arvanitaki:2009fg,Arvanitaki:2010sy,Yoshino:2012kn,Yoshino:2013ofa}.
The axion is a well-motivated dark matter candidate, though the distribution of such dark matter around the universe remains an unknown and interesting topic. Considering that in early universe evolution the density variations of both matter and dark matter occurred and acted as the seeds for cosmic structures today, there are great reasons to believe that axion distributes in cosmic space unevenly.

The axion was proposed as the pseudo-Nambu-Goldstone boson from the spontaneous breaking of a U(1) global symmetry, the Peccei-Quinn(PQ) symmetry, to solve the strong CP problem in QCD \cite{Peccei:1977hh,Peccei:1977ur}. It was later realized that there can be more Axion-like pseudo-Nambu-Goldstone bosons related to various global symmetry in UV theories \cite{Weinberg:1977ma,Wilczek:1977pj}. If this spontaneous breaking of global symmetry happened before or during inflation, then the axion field value can be considered homogenous across the observed Universe, except for small iso-curvature and density perturbations. In this case, the axion density variation is not generically large. If the symmetry was broken after inflation, then the axion field can only be correlated across the local horizon size. Later when the temperature dropped to the order of the axion mass $T\sim {m_{\va}}$, the axion began to oscillate at the bottom of the potential, which is acquired by non-perturbative effects such as instanton. At the same time axion topological defects decay. The oscillation is not coherent across the horizon.  Notice that during the universe evolution history, the axion mass also depends on the temperature of the universe.  These different horizon patches of the axion may have large density variations.

Some of the initial axion density variations grow under gravity, eventually collapse into denser objects and go through Bose-Einstein condensation \cite{Sikivie:2009qn,Iwazaki:2014wka, Tkachev:2014dpa,Eby:2017azn,Fairbairn:2017sil,Helfer:2016ljl,Vaquero:2018tib,Braaten:2015eeu,Visinelli:2017ooc, Krippendorf:2018tei,Guerra:2019srj} or even form black holes \cite{Boskovic:2018lkj, Sen:2018cjt}. Another type of axion objects is the axion clouds around the spinning black hole. It is produced from the black hole superradiance or simply gravitational accretion, which can induce black hole instability \cite{Boskovic:2018lkj}.  
The typical size and radius of the dense axion object depend on the mass of the axion, the decay constant $\Lambda_{\va}$, assumptions on the initial density fluctuations and the thermal dependence of the axion mass. The masses of such dense objects range from $10^{-18} M_{{\odot}}{\sim}10^{2} M_{{\odot}}$  \cite{Iwazaki:2014wka,Tkachev:2014dpa,Eby:2017azn,Fairbairn:2017sil,Helfer:2016ljl,Vaquero:2018tib}. The formation and evolution of such objects include some highly non-linear process and can be studied by numerical calculation. Furthermore, if the axion self-interaction is included or the density is very large, such axion dense objects can go through Bose-Einstein condensation and become even denser. Depending on the model, the average size can be approximated in some case around $10^2$ km for QCD axion, and could be possible candidates for the observed fast radio bursts (FRBs)~\cite{Iwazaki:2014wka,Tkachev:2014dpa,Eby:2017azn,Fairbairn:2017sil,Helfer:2016ljl,Vaquero:2018tib,Platts:2018hiy}. 
Moreover, there are various kinds of multi-messenger signals of the axion-like dark matter \cite{Graham:2015ouw,Liu:2019brz,Caputo:2019tms, Edwards:2019tzf, Jung:2020aem}, and the recent supermassive black hole image from the Event Horizon Telescope (EHT) can be used to probe the existence of ultralight bosonic particles that accumulate through the black hole superradiance effect~\cite{Akiyama:2019cqa, Chen:2019fsq}. 

%Bose-condensation %1/( {m_{\va}}{v_{\va}})\sim 

Here in this paper, we study a scenario that high energy bursts or periodic bursts of gravitational waves can be generated around the axion objects, including axion mini-clusters and axion clouds around the black holes, with the gravitational Chern-Simons ({CS}) couplings between axion like particle and gravitons \cite{Jackiw:2003pm,Alexander:2009tp}. It is well-studied that the electromagnetic radio bursts can be produced through similar processes, from the axion-photon couplings. We do a similar analysis with the gravitational {CS} coupling and the gravitational wave production from the drastic axion/photon energy oscillation and discuss their properties on the frequencies, strains, duration and the energy released. Such bursts have distinct signatures and interesting consequences and we can look for them with proposed different types of gravitational wave detectors  \cite{Arvanitaki:2012cn, Ito:2019wcb}. 

This paper is organized as follows.
In section \ref{PBH} we discuss the cases with the axion clouds around the primordial black holes. 
In section \ref{SecGWaxion}, we turn on both axion-photon and gravitational CS couplings and discuss different sources and gravitational strains produced.  We end with discussions and comments in section \ref{SecCon}. 
 In the appendix \ref{SecAxion} we review the FRB discussion for axion-photon couplings. In appendix \ref{SecGW} we calculate the gravitational waves induced by gravitational {CS} couplings on a flat background and discuss various features of such gravitational waves.

\iffalse
This paper is organized as follows.
% In Sec. \ref{SecAxion} we review the FRB discussion for axion-photon couplings. In Sec. \ref{SecGW} we calculate the gravitational waves induced by gravitational {CS} couplings on a flat background and discuss various features of such gravitational waves. 
In Sec. \ref{SecGWaxion} we turn on both axion-photon and gravitational CS couplings and discuss different sources and gravitational strains produced. We also discussed the cases with the axion clouds around the primordial black holes. We end with discussions and comments in Sec. \ref{SecCon}. 
\fi

\section{Axion clouds and primordial black holes}\label{PBH} 
\vspace{5pt}
%{ \it \bf Axion clouds and primordial black holes.}---
In this section, we show that for the axion with mass $\mu$eV (QCD axion) surrounding the black holes, the mass regime of the black holes that the axion clouds can form is between $10^{-7}M_{\odot}$ and $10^{-3}M_{\odot}$, supporting the parameters used in the previous sections. Rather than the stellar mass black holes, the masses of primordial black holes mostly fall into this mass region.
See e.g. \cite{Khlopov:2008qy,Belotsky:2018wph} for the formation of primordial black holes and recent constraints.

%as low as $10^{-20} M_{\odot}$. 

We define the typical scale of the black hole $R_{BH}= {G_N M_{BH}}/{c^2}$ and the wavelength of the axions $\lambda_{\va}= {\hbar}/{(m_{\va} c)}$.
Then define the dimension-less number
\begin{align} %=\frac{G_N M_{BH} m_{\va}}{\hbar c^3}
\alpha\equiv \frac{R_{BH}}{\lambda_{\va}}\simeq \left(\frac{M_{BH}}{M_{\odot}}\right)\left(\frac{m_{\va}}{10^{-10}\text{eV}}\right).
\end{align}
Based on the derivations in \cite{Arvanitaki:2009fg,Arvanitaki:2010sy,Yoshino:2012kn,Yoshino:2013ofa}, for the near extremal rotating black hole,
the formulation times of the axion clouds are estimated as 
\begin{align} 
\tau_{\va\uparrow} &\simeq 10^7 e^{1.84 \alpha} R_{BH} ,\quad \alpha\gg{1}, \label{tap}\\
\tau_{\va\downarrow} &\simeq 24\, \alpha^{-9} R_{BH},\qquad  \alpha\ll{1}.\label{tam}
\end{align}
which are required to be smaller than the age of the universe $\sim10^{10}$ years, rescaled to be comparable to the axion clouds formation time: %Eddington time 
\begin{align} 
\tau_{U} & \simeq 10^{23} \left(\frac{M_{\odot}}{M_{BH}}\right) R_{BH}. \label{tu}
\end{align}

%\tau_{E} & \simeq 2.5\times 10^{21} \left(\frac{M_{\odot}}{M_{BH}}\right) R_{BH}  . 

\begin{figure}[h]
%\centering
\raggedright
\includegraphics[scale=0.6]{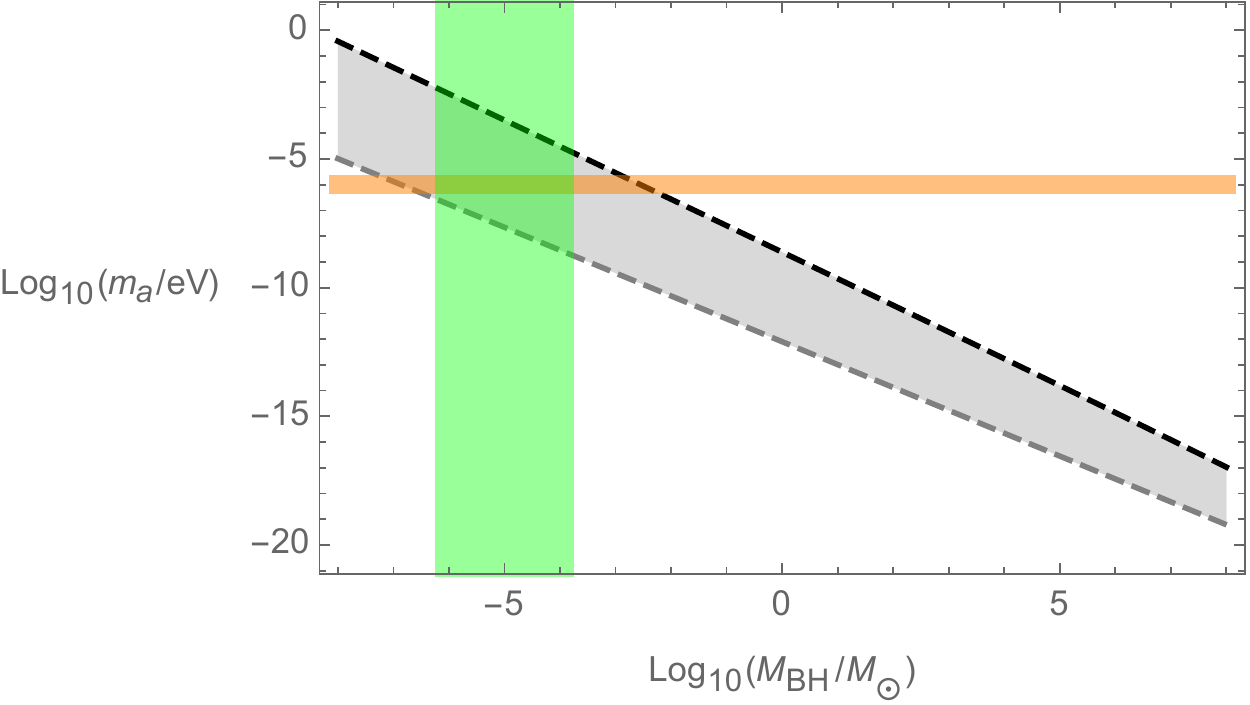}\qquad \\
\caption{The plot of axion mass $m_a$ and black hole mass $M_{BH}$ in the shaded area of light gray, which is bounded from $\tau_{\va\uparrow}<\tau_{U}$ and $\tau_{\va\downarrow}<\tau_{U}$ with \eqref{tap}\eqref{tam}\eqref{tu}. 
 The horizontal shaded region in light orange indicates the mass region of the axions around $\mu$eV.
The vertical shaded region in light green indicates the mass region of the black holes around $10^{-5}M_{\odot}$.
} \label{fig4}
\end{figure}
%\qquad
%\includegraphics[scale=0.5]{fig3b}
%Schematic diagrams. 

In Figure~\ref{fig4}, we plot the allowed mass range that the axion clouds can form for primordial black holes. 
In the shaded area, the upper bound is given by $\tau_{\va\uparrow}<\tau_{U}$, and the lower bound is given by $\tau_{\va\downarrow}<\tau_{U}$, with the expressions \eqref{tap}\eqref{tam}\eqref{tu}. 
Thus, for the clouds of axions with masses around $\mu$eV, the typical masses of host primordial black holes are around $10^{-5}M_{\odot}$, which in agreement with the early estimation in \cite{Rosa:2017ury}.
More intriguingly, it matches with the mass estimation of the novel object ``Planet 9" with a mass $5-10 M_{\oplus}$ in the outer Solar System \cite{Scholtz:2019csj,Witten:2020ifl}, which is just around the order of $10^{-5}M_{\odot}$.

%1.5-3 \times
%The high frequency gravitational waves from light primordial black holes can also see, e.g. \cite{Dolgov:2011cq}, and \cite{Nielsen:2019izz} for gravitational atom.  Recent progresses can also be found in \cite{Choi:2019mva,Hertzberg:2020hsz,Wang:2020zur,Chernoff:2020hdl}.

%\section{Interference of the EM and gravitational waves}
%\section{Joint analysis of the amplified EM and GW}

\section{Amplified electromagnetic waves and gravitational waves}
\label{SecGWaxion}  
%\vspace{5pt}
%{\it\bf Amplified electromagnetic waves and gravitational waves.} ---
In this section, we would like to discuss both the electromagnetic waves and gravitational waves as well as the interference effects coming from axion clumps. Especially, we consider the branching factor of the axion energy bursts into both waves and the gravitational waves created by the photon bursts. The total action of {CS} modified gravity with axion photon coupling is given by
\begin{align}
S_{\va}&= \int d^4 x\sqrt{-g}\Big( {  \frac{1}{2{\kappa_4}}} R -\frac{1}{4} F^2 +\L_{{\va}}+\L_{{\va}F\tilde{F}}+\L_{{\va} R\tilde{R}} \Big),
\end{align}
where $\L_{{\va}}$, $\L_{{\va} R\tilde{R}}$ and $\L_{{\va}F\tilde{F}}$ are given below:
\begin{align}
\L_{{\va}} &= -\frac{1}{2}(\partial{\va})^2- V({\va}).\label{thetaV}
\end{align}
\begin{align}\label{thetaFF}
 \L_{{\va} F\tilde{F}} &= -\frac{{\lf}}{4} {\va} F  \tF  \equiv - \frac{{\lf}}{4} {\va} F_{\mu\nu} \tF^{\mu\nu},
\end{align}
\begin{align}\label{thetaRR}
\L_{{\va} R\tilde{R}}& = \frac{{\lr}}{4} {\va} R\tilde{R} \equiv \frac{{\lr}}{4} {\va} R^\beta_{~\alpha\gamma\delta}
\tR^{\alpha~\gamma\delta}_{~\beta},
\end{align}
where $\tR^{\alpha~\gamma\delta}_{~\beta}\equiv\frac{1}{2}\epsilon^{\gamma\delta\mu\nu}R^\alpha_{~\beta\mu\nu}$ and $\tF^{\mu\nu}\equiv\frac{1}{2}\epsilon^{\mu\nu\lambda\rho}F_{\lambda\rho}$.

Again, we assume that the axion field ${\va}$ is only time-dependent. Except for the axion resonance, there are two other possible sources for the gravitational waves, although they are expected to be much smaller than the resonance. Those two sources are the stress energy tensors of the Maxwell field and axion field. 
The equation of the motion becomes:
\begin{align}
\label{GWFE2}
\Box h_{ij}  &=  {{\kappa_4}{\lr}} \tilde{\epsilon}^{p k}_{~~(i} \big[ \dot {\bar{\va}}  ( \partial_p \Box  h_{j)k})- \ddot {\bar{\va}}  (\partial_p \partial_t h_{j)k})  \big]\nonumber\\
&-2\kappa_4\left[T^{(\gamma)}_{ij} +T^{({\va})}_{ij} \right],
\end{align}
where $T_{\mu\nu}^{(\gamma)}= F_{\mu\alpha}F_{\nu}^{~\alpha}-\frac{1}{4}\big(F^2 + {\lf}{\va} F  \tF\big)g_{\mu\nu}$,
and $T_{\mu\nu}^{({\va})}= \partial_\mu{\va}\partial_\nu{\va}- \Big[ \frac{1}{2}(\partial{\va})^2+  V({\va}) \Big] g_{\mu\nu}$.
%$T^{(\gamma)}_{ij}$ and $T^{({\va})}_{ij} $ are given in 
%\begin{align}
%T_{\mu\nu}^{(\gamma)}&= F_{\mu\alpha}F_{\nu}^{~\alpha}-\frac{1}{4}\big(F^2 + {\lf}{\va} F  \tF\big)g_{\mu\nu},\label{Tgamma}\\
%T_{\mu\nu}^{({\va})}&= \partial_\mu{\va}\partial_\nu{\va}- \Big[ \frac{1}{2}(\partial{\va})^2+  V({\va}) \Big] g_{\mu\nu}.\label{Ttheta}
%\end{align}
At the meanwhile, the equations of motion for the electromagnetic field and axion field become 
\begin{align}
\nabla_\mu F^{\mu\nu} & =- {{\lf}}\partial_\mu{ {\va}} \tF^{\mu\nu}, \\ \label{Maxwell2}
(\Box- {m_{\va}^2}) {\va} &= \frac{{\lf}}{4} F \tF- \frac{{\lr}}{4} R \tR.
\end{align}
It requires involved numerical studies to solve the coupled equations, especially around the axion objects. 
%\end{align} The dynamic equation of motion for the axion field becomes 
%\begin{align}

\subsection{Estimation of the branching ratios}
%\vspace{5pt}
%{\it \bf Estimation of the branching ratios.} ---
In the following, we estimate the strength of both signals. One key property we would like to mention is the branching ratios from axion to both photons and gravitons, assuming both couplings turned on. Considering both photon and graviton are massless and have only two degrees of freedom, we have at the tree level at the zero temperature, 
\begin{align}\label{Br}
\frac{\text{Br}({\va} \rightarrow {\hg}{\hg})}{\text{Br}({\va} \rightarrow \gamma\gamma)}  \simeq  \frac{\lr^2}{\lf^2} .
\end{align}
Considering that the gravitational {CS} coupling ${\lr}$ is much less constrained than the photon-axion coupling ${\lf}$, we can expect that the energy of bursts can go to the gravitational sector dominantly.
 For example, if we take the typical values in 
$
{\lf}\simeq 10^{-2}/f_a \sim 10^{-14} \text{GeV}^{-1}.
$
and 
$
{\kappa_4}{\lr} =\frac{\ell^2}{M_P}< 10^8 \text{eV}^{-3},
 {\lr} =\frac{\ell^2 M_P}{2} < 10^{38} \text{eV}^{-1}.
$
which will lead to 
$\frac{\lr^2}{\lf^2}  \sim 10^{110}$. See the appendix for more discussion on the constraints of these couplings. That means the gravitational wave bursts associated with the {FRBs} can be tremendously large. Observing this will put a new constraint for $\lr$, apart from the current observation.

\begin{figure}[h]
\centering
%\raggedright
\includegraphics[scale=0.25]{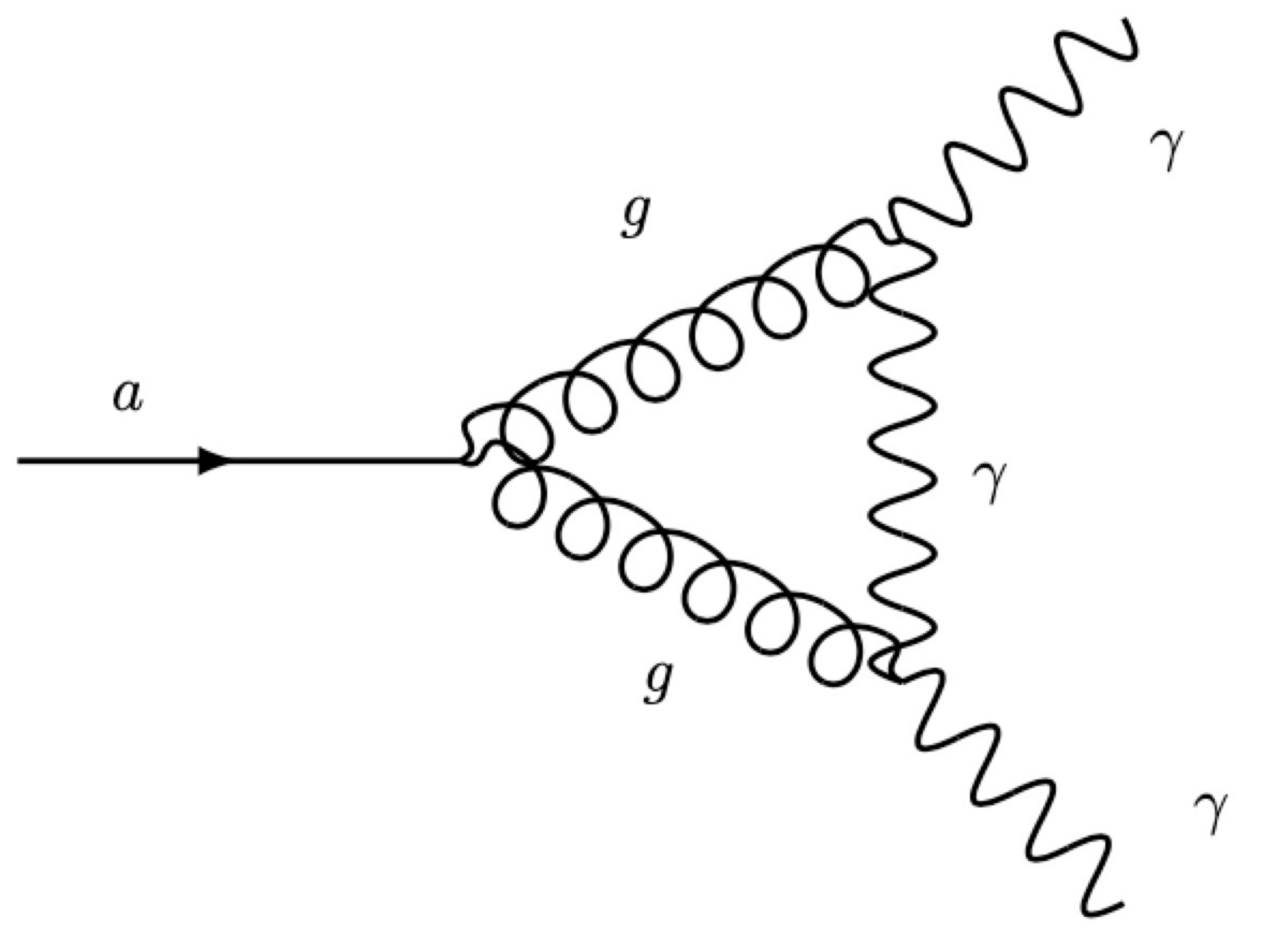}\\%{feynman1} \\{triangle2}
\includegraphics[scale=0.40]{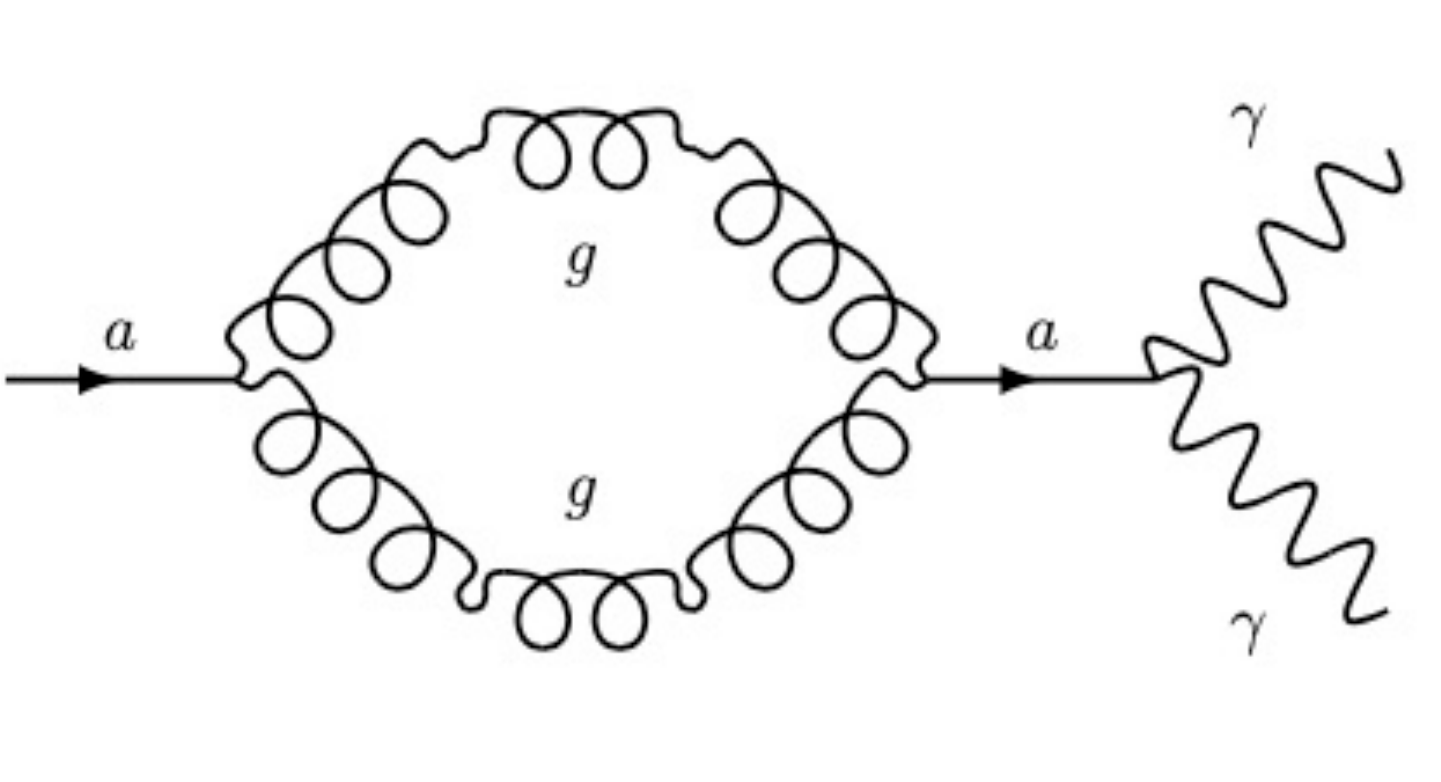}\\%{loop1}
\caption{
Top: The triangle Feynman diagram, where the axion-photon coupling is generated from axion-graviton coupling.
Bottom: The one loop diagram of graviton, which renormalizes the axion mass term, and provides the sub-leading correction to the axion-photon coupling.
} \label{figagg}
\end{figure}

It is interesting to discuss the CS gravity coupling a little more here. In UV-completed models, the CS gravity and photon-axion couplings are considered at the same order from compactification \cite{Becker:2005nb}. However in recent study \cite{Dyda:2012rj}, as well as our concern here, the CS gravity is treated as the low energy effective field theory with a cut-off. The existence of the ghost mode constrains the theory in certain parameter space, see \cite{Dyda:2012rj} for an interesting discussion.
 We refer to Fig.~\ref{figagg} to show some connections between two different kinds of CS couplings. This triangle diagram is divergent with 4 extra powers of momentum from $\va R\tilde{R}$ vertex and one power of momentum from each $h_{\mu\nu}T^{\mu\nu}$ coupling (see the interaction forms in \cite{S:2020zsp}), and evaluated as $\lf \sim \lr(\Lambda_c/M_{pl})^4$, where $\Lambda_c$ is the cut-off for Chern-Simons theory.  The cut-off scale $\Lambda_c$ is considered to be much lower than the Planck scale, which implies $ \lr/\lf \sim(M_{pl}/\Lambda_c)^4\gg 1$.
 %which might also be the reason why this CS gravity term hasn't been considered much for inflationary case.  
 Moreover, since the photon-axion coupling is a lot much constrained, this diagram can also be used for constraining the effect of the low energy Chern-Simon gravity theory. Besides, $\va R\tilde{R}$ also induces an axion self-energy loop with two gravitons running in the loop. This diagram is also divergent and renormalizes the axion mass term.  
 With a much lower cut-off scale $\Lambda_c\ll M_{pl}$, the one loop diagram provides the sub-leading correction to the axion-photon coupling. Here we only focus on the first triangle diagram, and assume that the axion-photon coupling is generated from axion-graviton coupling. 

For the $\va R\tilde{R}$ term, people have mostly studied its impact on the binary systems in the early inspiral phase, above the LISA sensitivities. The released power $P_{(\hg)}$ of the gravitational wave bursts can be related to the strain $h_{(\hg)}$ at the detector through $ h_{(\hg)}^2 \sim\frac{\kappa_4P_{(\hg)}}{{\rd}^2 \nu^2} $  \cite{Arvanitaki:2009fg,Arvanitaki:2010sy,Yoshino:2012kn,Yoshino:2013ofa}. 
From the {FRB} information, the power $P_{(\gamma)}$ is roughly $10^{42} $ ergs in $10^{-3} $ seconds
 % (from the redshift around $z  \sim 0.5$)  
\cite{Tkachev:2014dpa,Eby:2017azn,Fairbairn:2017sil,Helfer:2016ljl,Vaquero:2018tib}.
Assuming that $ {P_{(\hg)}}/{P_{(\gamma)}}\simeq  {\lr^2}/{\lf^2}  \simeq 10^{8}$, we have % {\cm we can calculate $h \sim ?$ at peak frequency.}millisecond
  \begin{align}\label{GWhg}
  h_{(\hg)} &\sim 10^{-23} \Big(\frac{1\text{GHz}}{\nu}\Big)  \Big(\frac{\lr/\lf}{10^4} \Big) \Big(\frac{1 \text{kpc}}{\rd}\Big).
     \end{align}
 %where $ P_{(\gamma)} \sim 10^{-12}M_{\odot}/ \text{ms}$.
%On the other hand, if we consider
The branching ratio in \eqref{Br} implies $\frac{P_{(\hg)}}{P_{(\gamma)}}\simeq\frac{\lr^2}{\lf^2}\gg 1$.
Then it is possible that $h_{(\hg)}$ can be greatly enhanced at the same frequency of photons $\nu={\ka}/2\pi={m_{\va}}/4\pi\sim$ GHz. 
%For example, if taking $\frac{\lr}{\lf}\simeq 10^{4}$, then $h_{(\hg)}\sim 10^{-23}$ or even higher can be reached.
Though we need to assume a much larger axion clump, instead of the size of mini-cluster estimated only from the power  $ P_{(\gamma)} \sim 10^{-12}M_{\odot}/ \text{ms}$ of the FRBs.

Now let us estimate the secondary effects of the gravitational waves productions from the photon burst process, with the source given by the stress energy tensor $T_{\mu\nu}^{(\gamma)}$ of the Maxwell field in \eqref{GWFE2}. The discussion of the stress-energy tensor of axion field $T_{\mu\nu}^{({\va})}$ in \eqref{GWFE2} is similar and generalization is straightforward. 
The quadrupole of the stress energy tensor is defined as $I_{ij}(t)= \int_{\ra} x_i x_j T_{tt}(t,\vec{x})d^3x$. The quadrupole integration is over the size of the axion object. And for the detector at the distance of ${\rd}$ from the source and assuming ${\rd}\gg\ra$, we have $h_{ij} (t)=\frac{\kappa_4}{4{\pi}{\rd}}  \ddot{I}_{ij}(t-{\rd})$. Notice that a sphere has zero quadruple moments and only out of the equilibrium decay process of the axion clump can induce the gravitational waves. We approximate $I_{ij} \sim \epsilon E_{(\gamma)} \ra^2$, where $E_{(\gamma)}$ is the total observed released energy of the photons in FRBs, and  $\epsilon E_{(\gamma)}$ is the non-spherical contribution.
Then $h_{ij} \sim \epsilon\frac{ {\pi}\kappa_4}{{\rd}} E_{(\gamma)} \ra^2 \nu^2 $, and we will compare this to the direct gravitational wave production from the {CS} gravitational term later. Considering $\epsilon\sim10^{-8}$, the strain $h_{(\va)}$ and $h_{(\gamma)}$ that our detector can receive are bounded by the dimensionless number 
\begin{align}\label{GWbound}
h_{(\gamma)}  \sim10^{-25} \Big(\frac{\nu}{1\text{GHz}}\Big)^2  \Big(\frac{E_{(\gamma)}}{10^{-12}M_{\odot}} \Big)\Big(\frac{1 \text{kpc}}{\rd}\Big) \Big(\frac{\ra}{10^2\text{km}}\Big)^2.
\end{align}

\begin{figure}[h]
%\centering
\includegraphics[scale=0.6]{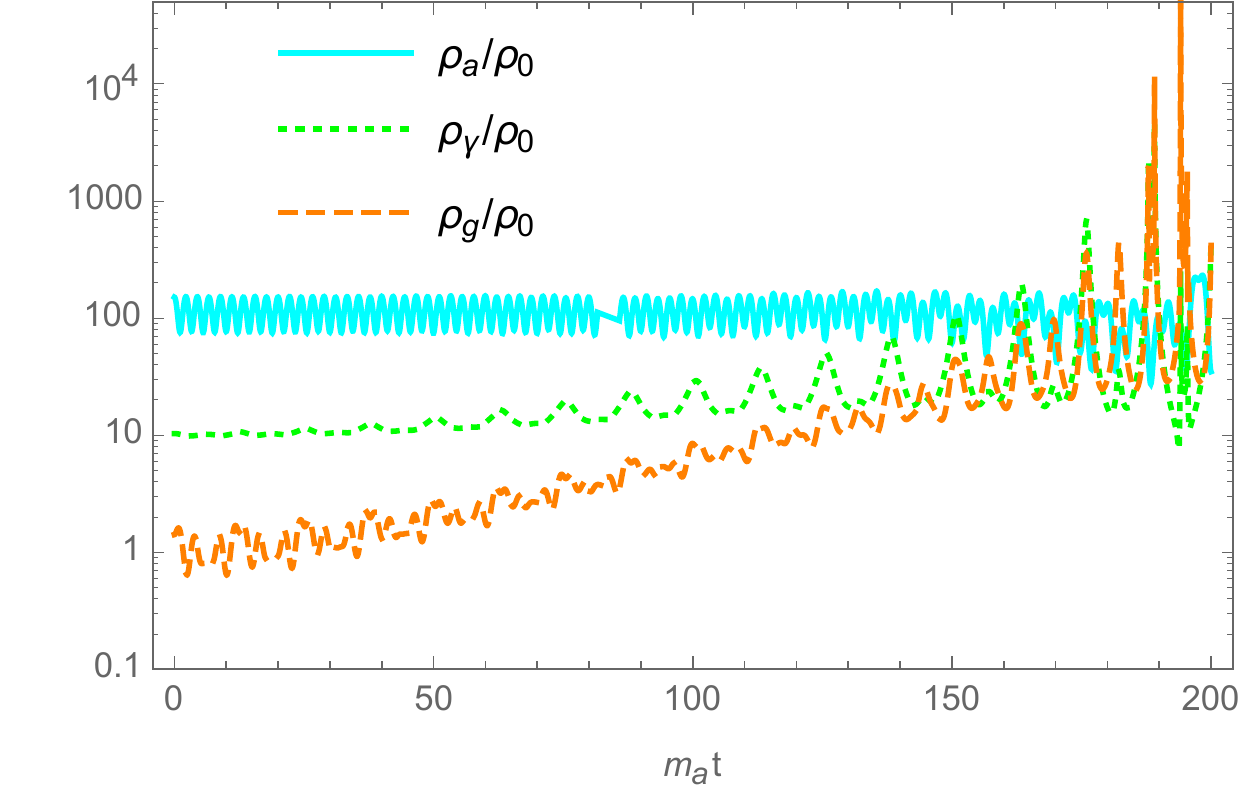} \\%\qquad\qquad
\includegraphics[scale=0.5]{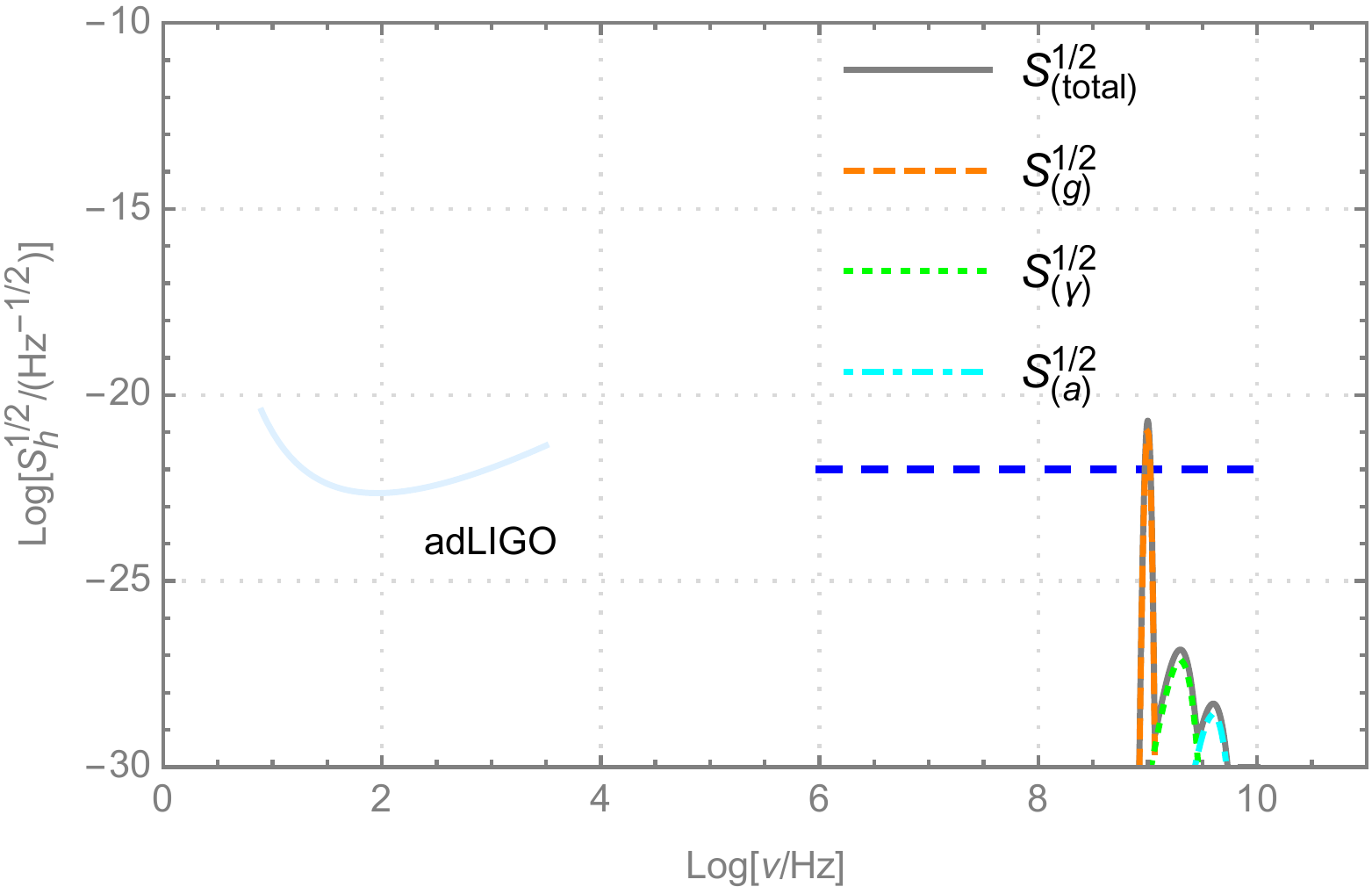}
\caption{
Schematic diagrams. 
Top: the growth of the energy density of axion $\rho_{\va}$, photons $\rho_\gamma$ and GWs $\rho_{\hg}$, which are normalized by the initial value $\rho_0\equiv \rho_g(t=0)$. We have chosen $\Gamma_{\hg}/{m_{\va}}=2\Gamma_{\gamma}/{m_{\va}} =1/40$  in the numerical study. 
%Bottom: the logarithmic plot of the strain $h$ in terms of typical frequency $k/m_{\va}\sim$GHz of different modes, where the rough values in  \eqref{GWhg} and \eqref{GWbound} have been taken, with $ {P_{(\hg)}}/{P_{(\gamma)}}\simeq 10^{8}$ and $L\simeq 1$Mpc. 
Bottom: the logarithmic plot of the strain sensitivity $S_h^{1/2}\equiv h_c/\nu^{1/2}$ in terms of typical frequency $\nu \sim $ GHz of different modes, where the rough values in  \eqref{GWhg} %and \eqref{GWbound}
 have been taken as $\lr/\lf \simeq 10^{10}$ and $L\simeq 1$kpc. 
 The light blue line is the expected strain sensitivity of advanced LIGO \cite{Ligo}, and the dashed blue line is based on the proposed sensitivities in \cite{Aggarwal:2020olq}
} \label{fig3}
\end{figure}

Another way to estimate the quadrupole moment is considering the non-spherical contribution $\ddot{I}_{ij}^{(\gamma)} \sim E_{n} v_{\gamma}^2$ and  $\ddot{I}_{ij}^{(a)}\sim E_{n} v_a^2$. 
By assuming $v_{\gamma} \sim c$,  we arrive $h_{(\va)}\sim h_{(\gamma)}  {v_a^2}/{c^2} $.
 Notice that the typical frequency $k$ are $m_a$ and $2m_a$ respectively for $h_{(\gamma)}$ and  $h_{(\va)}$, generated from different sources $T_{\mu\nu}^{(\gamma)}$  and $T_{\mu\nu}^{({\va})}$ with $v_a \sim 0.1$. 
%\begin{align}\label{GWha}h_{(\va)} \frac{c^2}{v_a^2} \sim h_{(\gamma)} \sim 10^{-28}   \Big(\frac{E_{(\gamma)}}{10^{-12}M_{\odot}} \Big)\Big(\frac{1 \text{kpc}}{\rd}\Big).
%\end{align}
We plot the schematic diagrams of these possible signals in Fig. \ref{fig3}.
In particular, we expect the signals around the spinning black hole is even stronger \cite{Boskovic:2018lkj} and may be within the range of the high frequency gravitational waves detection proposals \cite{Arvanitaki:2012cn,Ito:2019wcb}. When the size of the black hole is at the order of $1/(m_a v_a)$, we have the superradiance effects and the axion clouds form. For $m_a \sim$ GHz gravitational waves, the corresponding black hole mass is around $10^{-5} M_{\odot}$, and the bound of strains $h_{(\va)}$  becomes
\begin{align}\label{GWhg1}
h_{(\va)}\sim 10^{-27} \Big(\frac{1\text{GHz}}{\nu}\Big)  \Big(\frac{\epsilon_c M_\text{BH}/\text{ms}}{ 10^{-9} M_{\odot}/ \text{ms}} \Big)^{1/2} \Big(\frac{1 \text{kpc}}{\rd}\Big) ,
     \end{align}
where $\epsilon_c$ is the portion of the energy stored in the black hole clouds, and can be taken as $10^{-4}$. It is foreseeable that CS term corrects the black hole superradiance effects by some factor, e.g. see \cite{Alexander:2009tp}, or induce Kerr black hole instability \cite{Gao:2018acg}, although the overall order should not change much.
%
% \eqref{GWhg} \eqref{GWha} 

%
%which can match the order estimated in \eqref{GWhg1}. 

%The back-reaction is not considered here, the same as in Fig.1.

 One can also see more discussions on $h_{(\va)}$ in a black hole background in \cite{Arvanitaki:2012cn,Ito:2019wcb}, corresponding to $\va\,\va \rightarrow g$ annihilation. Also, notice here the estimations for the stain $h$ are very rough, since the collapsing and bursting processes are actually very complicated. The more precise estimation will require some involved numerical studies. From the estimations above and related pulsars or supernovae burst processes informations, these secondary gravitational effects without the {CS} gravity terms can release around $1\%$ of total released energy. 

%We will estimate the strain from direct gravitational wave production from the FRB-like sources in the next section.

Above the strains are estimated for GHz fast gravitational wave bursts. We expect that $h_{(\va)} \sim h_{(\gamma)}$ can be different from $h_{(\hg)}$ for MHz gravitational waves, and notice that the normal drastic FRB energy oscillations induce gravitational bursts $h_{(\va)} \sim h_{(\gamma)}$ as well without the {CS} gravitational coupling. 
 %For the late universe, previously the gravitational waves produced from  $\va F\tilde{F}$ or $\va R\tilde{R}$ have also been discussed in different settings. 
 For $\va F\tilde{F}$ the particle decay process $\va \rightarrow \gamma\gamma $ or $\va\va \rightarrow \gamma\gamma$ gives rise to a gravitational wave signal directly, although the signal is very small without the presence of a black hole. The authors of \cite{Arvanitaki:2012cn,Ito:2019wcb} discussed this kind of gravitational wave production enhanced around the black hole. Especially for the black hole superradiance the process $\va_+\rightarrow \va_- + {\hg}$ happens, where a superradiant cloud axion emits a graviton and jumps onto a lower level, then may get captured by the black hole horizon. A benchmark frequency of this process can be $10^{-2}$ Hz for a $10^7 M_{\odot}$ black hole and a $10^{-17}$ eV axion. And this could possibly be detected by LISA. Considering the distance between the detector and the source $L$, the strain is given by
$ \sim10^{-22} \big(\frac{10^{-2}\text{Hz}}{\nu}\big)^{1/2}\big(\frac{M_{\text{BH}}}{10^7 M_{\odot}} \big)^{1/2} \big(\frac{100 \text{Mpc}}{\rd}\big)$  \cite{Arvanitaki:2009fg,Arvanitaki:2010sy,Yoshino:2012kn,Yoshino:2013ofa} .

%In our parameter regime, 
 % \begin{align}\label{GWhs}
% $ h_{(s)} \sim 10^{-27} \Big(\frac{1\text{GHz}}{\nu}\Big)^{1/2}\Big(\frac{M_{\text{BH}}}{10^{-3} M_{\odot}}\Big)^{1/2} \Big(\frac{1\text{kpc}}{\rd}\Big)$.
%   \end{align}
%The order matches that in \eqref{GWhg1} and \eqref{GWha}.

%\vspace{10pt}
\section{Discussion and Summary}
\label{SecCon} 
\vspace{5pt}
%{\it\bf Discussion and Summary.} ---
We have studied the gravitational wave bursts from the {CS} coupling with axion and graviton as well as the associated FGB from FRB sources without the {CS} term, though we only performed the calculation on the flat background to demonstrate the features of the signals. For the case with axion clouds around the spinning black hole, the situation is more complicated. Especially the black hole superradiance can keep on copiously producing axions, and induce the instability around the axion clouds, where the gravitational bursts/radio bursts can happen periodically.

We would also like to comment on the fact that the CS gravity has a ghost in the UV, which can be fixed by higher order terms, e.g.~\cite{Qiao:2019wsh,Zhao:2019xmm,Zhao:2019szi}. Moreover, our discussions, even around the black holes, are in the IR region, and it is perturbative in the gravitational release case.
The parametric resonance effects of both electromagnetic and gravitational fields are important ingredients. Especially the gravitational wave bursts can be dominant in the energy release. In that case, the radio burst signals can be a reference point for the gravitational releases. Notice that the CS gravity coupling can be much larger than the axion-photon coupling, due to the limited studies and experiments on the bound.

For the cosmological inflation with axion term, the GW signal/photon signal is usually circularly polarized, since one mode is growing and the other mode is decaying. This is due to the fact that the background field $\frac{d\theta}{dt}$ is with the fixed sign. However, in our case with the oscillating background $\theta \propto \sin(m t)$, both of modes are growing exponentially. So the signals do not have a certain polarization. The oscillating background only modifies the phase factor of the signal modes, e.g. see  \cite{Yuan:2020xui}  for a related discussion. 

In summary, the gravitational wave burst signals proposed here happened in the late universe and can be strongly enhanced by the gravitational {CS} couplings inside the dense axion regions. Its characteristics include high frequencies and burst features within a short period. For this process around the spinning black holes, we expect some technical issues to overcome~\cite{Loutrel:2018ydv}. The spinning black hole solution is corrected in CS gravity. And the black hole accretion process and the axion cloud dynamics are modified from this enhanced decays of $\va \rightarrow{\hg\hg}$. Detecting the gravitational waves of high frequencies around kHz to GHz is also an interesting topic for future detectors.
In addition, there could be entangled signals from particles decays, which might be detected with the HBT interferometers, as suggested in \cite{Chen:2017cgw, Kanno:2018cuk}.
 For the discussion on gravitational waves from light primordial black holes one can refer to \cite{Dolgov:2011cq}, and recent progresses on the FRBs and resonant instability as well as superradiance around primordial black holes can also be found in \cite{Nielsen:2019izz, Choi:2019mva,Hertzberg:2020hsz,Wang:2020zur,Chernoff:2020hdl,Arza:2020eik,Buckley:2020fmh,Ferraz:2020zgi}.

%
%We would mention that around the strong gravitational background, axion can convert to graviton directly assuming {CS} couplings. This effect is similar to the axion-photon conversion in the presence of the magnetic field and can enhance the gravitational bursts discussed above. 
%The detailed study of such effect is left for a future work. 

\appendix
\section{Axion dense objects and fast radio bursts }
\label{SecAxion}

Let us consider the model of electromagnetic and axion field with an axion photon interaction term,
\begin{align}\label{EM1}
S_{EM}&= \int d^4 x\sqrt{-g}\left( -\frac{1}{4} F^2+\L_{\va} +\L_{{\va} F\tilde{F}}\right),
\end{align}
where $F^2\equiv F_{\mu\nu} F^{\mu\nu}$ and the electromagnetic tensor $F_{\mu\nu}=\partial_\mu A_\nu-\partial_\nu A_\mu$. The Lagrangian density of the axion field ${\va}$ is given by
\begin{align}
\L_{{\va}} &= -\frac{1}{2}(\partial{\va})^2- V({\va}).\label{thetaV}
\end{align}
The usual choice of the potential is $V({\va})={m_{\va}^2}{f_{\va}^2}\big(1-\cos\frac{\va}{f_\va}\big)$, 
where ${m_{\va}}$ is the axion mass and $f_{\va}$ is the axion decay constant.
We are interested in the region $\va\ll{f_\va}$, such that the potential in \eqref{thetaV} is approximately $V({\va})\simeq \frac{1}{2}{m_{\va}}{\va}^2$.
%
%The coupling ${\gp}$ and ${{\lf}}$ has a momentum dimension $-1$,  those two are all dimension-5 operators with axion and dilaton couplings. 
The interaction term between the axion and Maxwell field is
\begin{align}\label{thetaFF}
 \L_{{\va} F\tilde{F}} &= -\frac{{\lf}}{4} {\va} F  \tF  \equiv - \frac{{\lf}}{4} {\va} F_{\mu\nu} \tF^{\mu\nu},
\end{align}
where $\tF^{\mu\nu}\equiv\frac{1}{2}\epsilon^{\mu\nu\lambda\rho}F_{\lambda\rho}$.

%The coupling ${{\lf}}$ has a momentum dimension $-1$ and it is a dimension-5 operator. % with axion coupling.
The current experiments are now probing the parameter region $10^{-16} \text{GeV}^{-1}\lesssim{{\lf}}  \lesssim 10^{-8} \text{GeV}^{-1}$ and ${m_{\va}} \lesssim 10^4 \text{ eV}$ and part of the region has already been ruled out \cite{Arvanitaki:2009fg,Graham:2015ouw}. For the QCD axion, we have to impose the condition ${f_\va}{m_{\va}} = f_\pi m_\pi$, where $f_\pi$ and $m_\pi$ are the decay constant and mass of the pion, respectively. 
However, for general axion-like particles, the relation between ${f_\va}$ and ${m_{\va}}$ is model dependent and we consider two parameters as independent of each others. For a model of largely homogenous axion vacuum dark matter and considering that the oscillating axion accounts for no more than all dark matter, we have the relic abundant constraint ${f_\va}\lesssim {\ci}^{-2} \times10^{12} \text{GeV}$ with ${\ci} \sim \mathcal{O}(1)$ for initial misalignment. 
\begin{align}\label{lambdaF}
{\lf}\simeq 10^{-2}/f_a \sim 10^{-14} \text{GeV}^{-1}.
\end{align}

Consider the flat background $ds^{2} =\eta_{\mu\nu} d x^\mu d x^\nu$, where $\eta_{\mu\nu}=\text{diag}{[-1,1,1,1]}$.
From the action in \eqref{EM1}, the equations of motion for the axion field and Maxwell field are
\begin{align}
(\partial_\mu \partial^\mu - {m_{\va}^2}) {\va}  &=  \frac{{\lf}}{4} F \tF, \label{axion1} \\ 
\partial_\mu F^{\mu\nu} &=- {{\lf}}\tF^{\mu\nu}\partial_\mu{ {\va}}   . \label{Maxwell1}
\end{align}
We consider the case that the effects from axion coupling term is very small, such that the right-hand side of \eqref{axion1} can be ignored at the background level. Then we assume a coherent oscillation of the axion field as below,
%. If we assume an oscillation profile for now as below, and ignore the spacial dependence, 
\begin{align} \label{sin}
\langle \va\rangle= {\bar{\va}}(t)={\vt_0} \sin  { ( {m_{\va}} t+\phi_0)},
\end{align}
where ${\vt_0}\equiv{f_\va}\theta_i\, $.
Without loss of generality, we will adopt the phase $\phi_0=0$.

Note that for now, we do not consider the spatial dependence of the axion field value for simplicity, although the axion field today is assumed to be incoherent across space. It means that any large scale resonance is nearly impossible.  The incoherence is due to the realization of the axion oscillation in the early universe, and the axions field at the different spatial points may pick up different phases factors.  However,  the local axion object can play an important role in this electromagnetic amplification effect and has been discussed extensively, see e.g. \cite{Iwazaki:2014wka,Tkachev:2014dpa,Eby:2017azn,Fairbairn:2017sil,Helfer:2016ljl,Vaquero:2018tib}. Thus, we consider the coherent solution \eqref{sin} above as the background solution.

We assume that the gauge potential $A_\mu(t,z)$ is homogeneous in $x$- and $y$-directions and define the helicity $\pm$ modes $A_\pm(t,z)\equiv [A_x(t,z)\pm iA_y(t,z)]/\sqrt{2}$.
In the Coulomb gauge, the equations of motion for the Maxwell field \eqref{Maxwell1} are reduced to the formula
\begin{align}
\big[ -\partial_t^2+  \partial_z^2   \mp i   {{\lf}} \dot{ \bar{\va}}(t) \partial_z \big] {A}_\pm(t,z)=0 \,. \label{eomEM}
\end{align}
%\begin{align}
%A_\mu=a_x e_x+a_y e_y=a_+ e_-+a_- e_+ =e^{i\theta} e_-+ e^{- i\theta} e_+
%\end{align}
%\iffalse\fi
%We make an ansatz about the Maxwell field with typical wave number $k$ and
%\int\frac{d k}{\sqrt{2\pi}}\left[\right]
With the ansatz of the axion field in \eqref{sin}, $\dot{\bar{\va}}(t)$ is a cosine type function.
Consider the following mode, %Fourier transform 
\begin{align}
A_\pm(t,z)=  {b}_\pm(t)e^{  i k  z} . %+\text{h.c.\,} . ,\quad {b}_-(t)={b}_+^{\dagger}(t)
\end{align}
In the momentum space, the equation of motion \eqref{eomEM} is reformulated as
\begin{align} 
\ddot{b}_\pm(t)+  \Big[k^2 \mp   {{\lf} k}\dot{\bar{\va}}\Big] {b}_\pm(t)=0\,. \label{eom1}
\end{align}
%This equation \eqref{eom1} is nothing but the Mathieu equation $\ddot{y}(t)+ [b\pm 2q \text{cos}(2t)]y(t)=0$ for the parametric resonance. 
%For example, if we have Mathieu's differential equation, 
%The solution is the linear combination of the Mathieu cosine and Mathieu sine functions. We can write the solution as $y_\pm = e^{\pm i \mu_k  t } P_\pm (b, q, t)$, where $\mu_k$ is the complex Floquet exponent, $P_\pm$ is the periodic function of $t$. 
Redefine ${\tau}\equiv {m_{\va}} t/2$, we can write down \eqref{eom1} as 
$\frac{\partial^2{b}_\pm({\tau})}{\partial{\tau}^2} +\Big[ \frac{4k^2}{{m_{\va}^2}} \mp  {{\lf}}{\vt_0} \frac{4k}{{m_{\va}}} \cos(2 {\tau})  \Big] {b}_\pm( {\tau})=0\,$. 
It is nothing but the Mathieu equation $\ddot{b}(\tau)+ [p\mp 2q \text{cos}(2\tau)]b(\tau)=0$ for the parametric resonance
with $p=\frac{4k^2}{m_{\va}^2}, q= {{\lf}}{\vt_0} \frac{2k}{m_{\va}}$.
The particular solutions of \eqref{eom1} are then given by ${b}_\pm(t) \propto e^{  i \mu_k t } P_{\mp}\big(\frac{4k^2}{{m_{\va}^2}}, {{\lf}} {\vt_0}  \frac{2k}{{m_{\va}}}, \frac{{m_{\va}} t}{2}\big)$, where $\mu_k\simeq\frac{m_{\va}}{2}(1\mp i  \frac{{{\lf}} {\vt_0}}{2} )$.
The resonance can be achieved when $\va$ decays to the diphoton, at the resonance frequency ${{\ka}}={m_{\va}}/2$.
And a typical amplification happens when $\mu_k t > 1$.

Consider the size of the axion stars or mini-cluster with average size $\ra$ as the light travels through, 
then the total time $t_\gamma\simeq \ra /c\simeq   {1}/{({m_{\va}} {v_{\va}}c)}$, where ${v_{\va}}$ is the typical velocity inside of the clumps and the typical relation $\ra\simeq {1}/({m_{\va}} {v_{\va}})$ where the equilibrium cluster has been used  \cite{Iwazaki:2014wka,Tkachev:2014dpa,Eby:2017azn,Fairbairn:2017sil,Helfer:2016ljl,Vaquero:2018tib}. Thus, the amplification factor for the Maxwell field is $e^{\Gamma_{\gamma} t_\gamma}$, with
\begin{align}\label{GammaF} 
\Gamma_{\gamma}=   {{{\lf}} {\vt_0}}  \frac{{m_{\va}}}{2},\quad t_{\gamma} \simeq \frac{1}{{m_{\va}} {v_{\va}}c}.
\end{align}
We can see that the explosive decay of the axion clumps or axion clouds can happen, with certain axion parameters. 

\begin{figure}[h]
%\centering
\quad\includegraphics[scale=0.48]{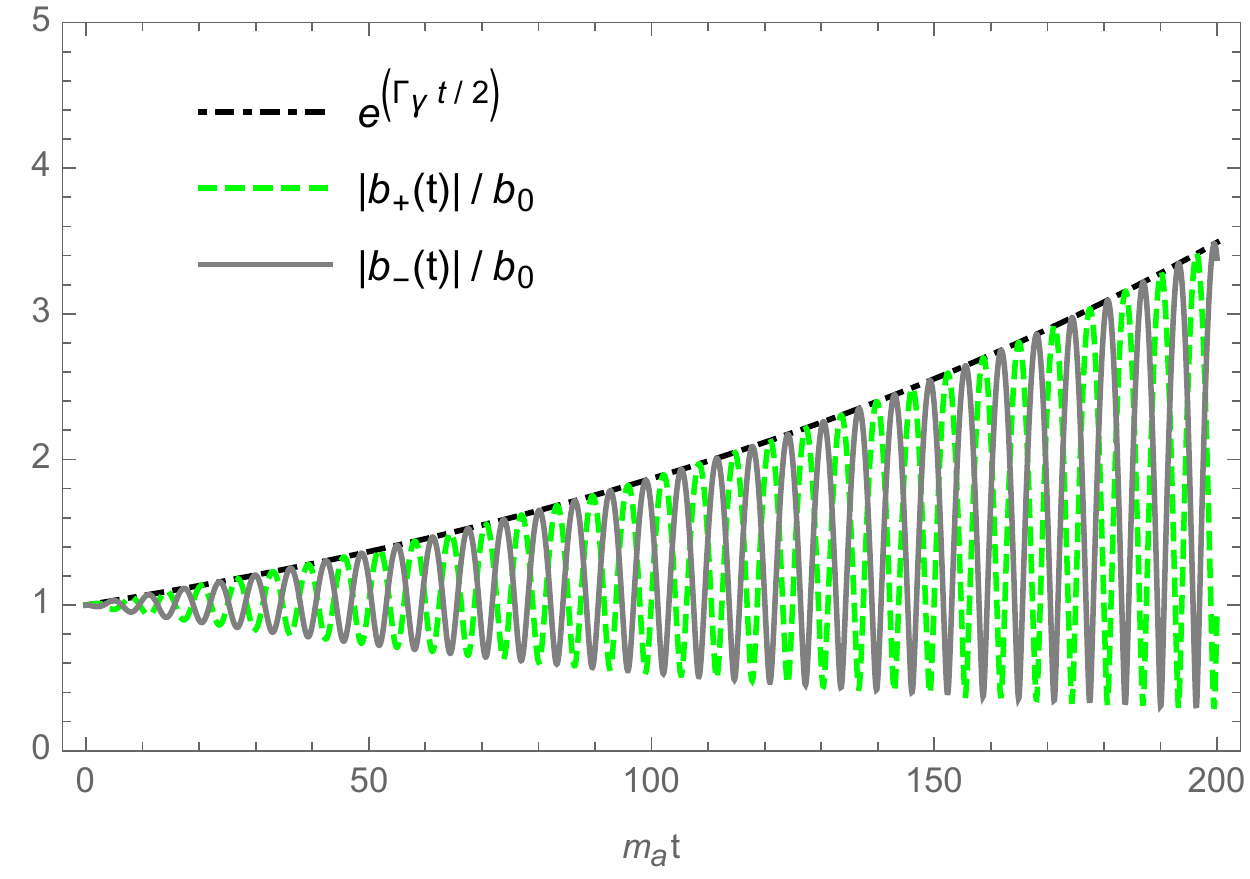}\\%\qquad
\includegraphics[scale=0.5]{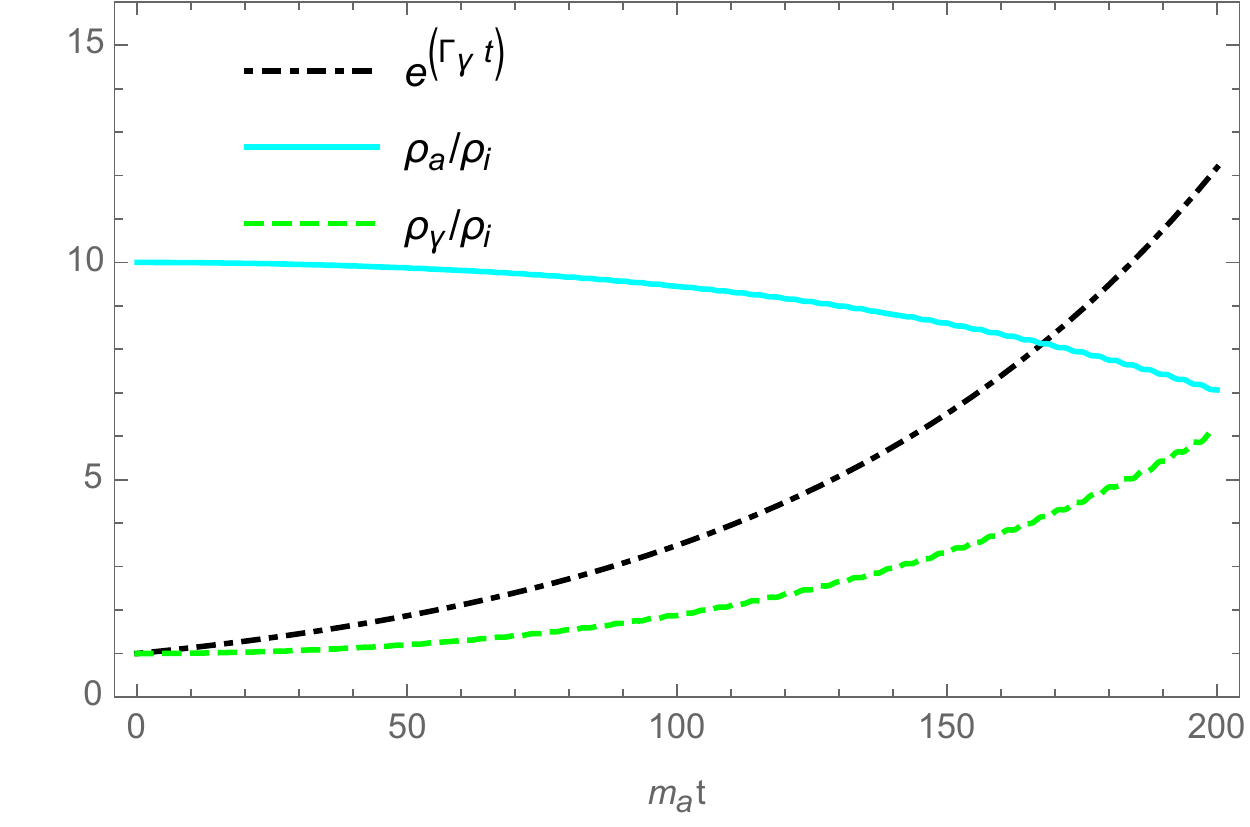}
\caption{
Schematic diagrams of the amplification of electromagnetic wave, where $\Gamma_{\gamma}/{m_{\va}}= 1/80$ in \eqref{GammaF} has been chosen in the numerical. 
Top: the amplitude growth of the helicity $\pm$ modes of gauge potential factors $b_\pm$ at the typical wave number ${\ka}={m_{\va}}/2$ in \eqref{eom1}, which are normalized by the initial value ${b_0}$. 
Bottom: the growth of energy density of the photons $\rho_\gamma$ and axion $\rho_{\va}$, which are normalized by the initial value of $\rho_{i}\equiv \rho_\gamma(t=0)$. Notice here we do not consider the back-reaction of EM and axion fields, but $\rho_a$ decreases slightly along with the amplification of $\rho_\gamma$ . } \label{fig1}
\end{figure}

At the meanwhile, we have the dynamic equation of motion for the axion field \eqref{axion1}. The complete solution of both axion and electromagnetic fields dynamics requires numerical studies. For example, see \cite{Boskovic:2018lkj} for the curved background solution.  Here we only present the qualitative description of the process that axion oscillation induces a burst of energy in the electromagnetic sector at the resonance frequency and the axion fields lost energy. With the ansatz of the axion field with the typical wave number ${\ka}$
\begin{align}\label{thetak}
\va(t,z) ={\bar{\va}}(t)+\delta \va(t, z), %{\delta\va}(t)+ {\va_k}(t)e^{ 2i {\ka}  z}+\text{h.c.\,},
\end{align}
the spacial factor in \eqref{axion1}  can be decomposed. It is enough to see a couple of features in this effect to explain the {FRBs}. 
In Figure \ref{fig1}, we plot the schematic diagrams of the amplification of electromagnetic waves. The energy density of the photons $\rho_\gamma=T_{tt}^{(\gamma)}$ and axions $\rho_{\va}=T_{tt}^{({\va})}$ are from the stress energy tensors
\begin{align}
T_{\mu\nu}^{(\gamma)}&= F_{\mu\alpha}F_{\nu}^{~\alpha}-\frac{1}{4}\big(F^2 + {\lf}{\va} F  \tF\big)g_{\mu\nu},\\
T_{\mu\nu}^{({\va})}&= \partial_\mu{\va}\partial_\nu{\va}- \Big[ \frac{1}{2}(\partial{\va})^2+  V({\va}) \Big] g_{\mu\nu}.
\end{align}

When the Maxwell fields pass through these axion dense objects or the axion clouds, due to the photon-axion interaction, the coherent oscillation of axions amplifies the Maxwell field. The result is an explosion of the photon number in the resonance frequency $ {{\ka}}={m_{\va}}/2$,
as in the mechanism $\va\to \gamma\gamma$ described in the introduction.  This effect has been amazingly linked to the observed multiple {FRBs} in the galaxy, in both the axion dense object case \cite{Iwazaki:2014wka,Tkachev:2014dpa,Eby:2017azn,Fairbairn:2017sil,Helfer:2016ljl,Vaquero:2018tib} and black hole superradiance \cite{Boskovic:2018lkj, Edwards:2019tzf,Chen:2019fsq}. Especially, it is for the peak frequency, 
\begin{align}
 &{{\ka}}={m_{\va}}/2, \quad  10^{-6} \mu\text{eV}<{m_{\va}} < 10^{4} \mu\text{eV}, 
\end{align} 
with a benchmark point frequency around GHz($\sim \mu$\text{eV}). 
%Liu:2019brz,Caputo:2019tms,
%10^{9} Hz =
%{\cm YL: either using ``frequency" or ``frequency", please check eq.(29) in Jiro's 1708.09592 \cite{Yoshida:2017cjl}}

\section{Axion amplified gravitational waves}
\label{SecGW}

In this section, we discuss the similar amplification effects of gravitational waves in the dynamical {CS} modified gravity \cite{Jackiw:2003pm, Alexander:2009tp}. 
%
%and postpone any discussion of GW generation to the next section. 
%Moreover, we shall not discuss in this section cosmological power spectra, since these will be summarized in Sec.~\ref{SEC:CScosmology}. 
%GW propagation in CS modified gravity has only been studied in the non-dynamical formalism with $\beta = 0$ and $\alpha = {\kappa_4}$. Let us begin with a discussion of GW propagation in a FRW background. 
%of dynamical {CS} modified gravity
The total action is given by
\begin{align}
S_{GW}&= \int d^4 x\sqrt{-g}\left( \frac{1}{2{\kappa_4}} R  + \L_{{\va}} +\L_{{\va} R\tilde{R}} \right),
\end{align}
where ${\kappa_4}=\frac{8\pi G}{c^4}=\frac{1}{M_P^2}\frac{\hbar}{c^3}$. The Lagrange density of the axion field  $ \L_{{\va}}$ is the same as in \eqref{thetaV}, and the {CS} term is given by
\begin{align}\label{thetaRR}
\L_{{\va} R\tilde{R}}& = \frac{{\lr}}{4} {\va} R\tilde{R} \equiv \frac{{\lr}}{4} {\va} R^\beta_{~\alpha\gamma\delta}
\tR^{\alpha~\gamma\delta}_{~\beta},
\end{align}
where $\tR^{\alpha~\gamma\delta}_{~\beta}\equiv\frac{1}{2}\epsilon^{\gamma\delta\mu\nu}R^\alpha_{~\beta\mu\nu}$.
% \L_{{\va} F\tilde{F}} &= -\frac{{\lf}}{4} {\va} F  \tF  \equiv - \frac{{\lf}}{4} {\va} F_{ab} \tF^{ab}, 
%\qquad \tF^{ab}\equiv\frac{1}{2}\epsilon^{abcd}F_{cd}, \\
%\L_{{\va}} &=  \int d^4 x\sqrt{-g}\big[-\frac{1}{2}(\partial{\va})^2-V({\va})\big],\qquad    V({\va})\simeq \frac{1}{2}{m_{\va}^2}  {\va}^2. %?

We consider the flat background with the gravitational wave perturbation $h_{ij}$,
\begin{align}
ds^{2} =  - d t^{2} + \left( \delta_{ij} + h_{ij} \right) d x^{i} dx^{j} . %a^{2}(\eta) \left[\right],
\label{FRW-metric}
\end{align}
And we take the transverse and traceless (TT) gauge, $ h^{i}{}_{i} = \delta^{ij} h_{ij} = 0$ and $\partial_{i} h^{ij} = 0$. 
Assuming that ${\va}$ is only time-dependent as in \eqref{sin}, the linearized Einstein field equations are given by
\cite{Alexander:2004wk,Alexander:2007kv, Alexander:2009tp,Yoshida:2017cjl,Yoshida:2017ehj},
\begin{align}
\label{GWFE}
%\Box h^{i}{}_{j}  = {{\kappa_4}{\lr}} \tilde{\epsilon}^{p k (i}
\Box h_{ij}  = {{\kappa_4}{\lr}} \tilde{\epsilon}^{p k}_{~~(i}
 \big[ \dot {\bar{\va}}  ( \partial_p \Box  h_{j)k})- \ddot {\bar{\va}}  (\partial_p \partial_t h_{j)k})  \big],
\end{align}
where $\Box=\nabla_\mu\nabla^\mu$ and $\tilde{\epsilon}^{p k i}= {\epsilon}^{t p k i}$.
%$ {{\kappa_4}{\lr}}\equiv \frac{{\lr}}{M_P^2}$.
%where $\bar{\square}$ is the D'Alembertian operator associated with the background.
%where the amplitude $h_{R}, h_{L}$ %, the unit vector in the direction of wave propagation $n_k$ 
%and the conformal wavenumber ${{\ka}}>0$ are all constant. 
%It is convenient to decompose into definite parity states, such as
%The amplitude is decomposed as
%\be \mathcal{A}_{ij} = h_{R}\e^{R}_{ij} +h_{L}\e^{L}_{ij} \ee
%$\e^{R,L}_{ij}$ are given in terms of the linear ones $\e^{+,\times}_{ij}$ by
%~\cite{Misner:1973cw} 
%\ba

Let us now concentrate on gravitational wave perturbations, for which one can make the ansatz
\begin{align}\label{hijmetric}
h_{ij}(t,z) =  \big[ h_{R}(t) \e^{R}_{ij} +h_{L}(t)  \e^{L}_{ij} \big]e^{ i {k} z}.  %+\text{h.c.\,},
\end{align}
The circular polarization tensors are defined as
$\e_{kl}^{R} =  \frac{1}{\sqrt{2}}\left(\e^{+}_{kl} + i\e^{\times}_{kl}\right)$ and 
$\e_{kl}^{L}  =  \frac{1}{\sqrt{2}}\left(\e^{+}_{kl} - i\e^{\times}_{kl}\right)$.
The modes $h_R=\frac{1}{\sqrt{2}}(h_+  - i h_{\times} )$ and $h_L =\frac{1}{\sqrt{2}}(h_+  + i h_{\times} )$.
These polarization tensors satisfy the condition $ \epsilon^{z jk} \e_{kl}^{\text{I}} = i
\varepsilon_{\text{I}} \left(\e^j{}_l\right)^{\text{I}}$, where ${\eR} = +1$ and ${\eL} =-1$. 
Then in the momentum space with $I=R, L$, \eqref{GWFE} becomes 
\begin{align} \label{hit}
&\big[ \ddot{h}_I(t)+ k^2 {h}_I(t) \big]\big[ 1-  \ve_I {{\kappa_4}{\lr}}k \dot{\va}(t) \big] 
=   {\ve_I{{\kappa_4}{\lr}}k \ddot{\va}(t)}  \dot{h}_I(t).
\end{align}
At the meanwhile, the dynamical equation of motion for the axion field is
\begin{align}
&(\Box - {m_{\va}^2}) {\va} = - \frac{{\lr}}{4} R \tR.
\end{align}

\begin{figure}[h]
%\centering
\quad\includegraphics[scale=0.48]{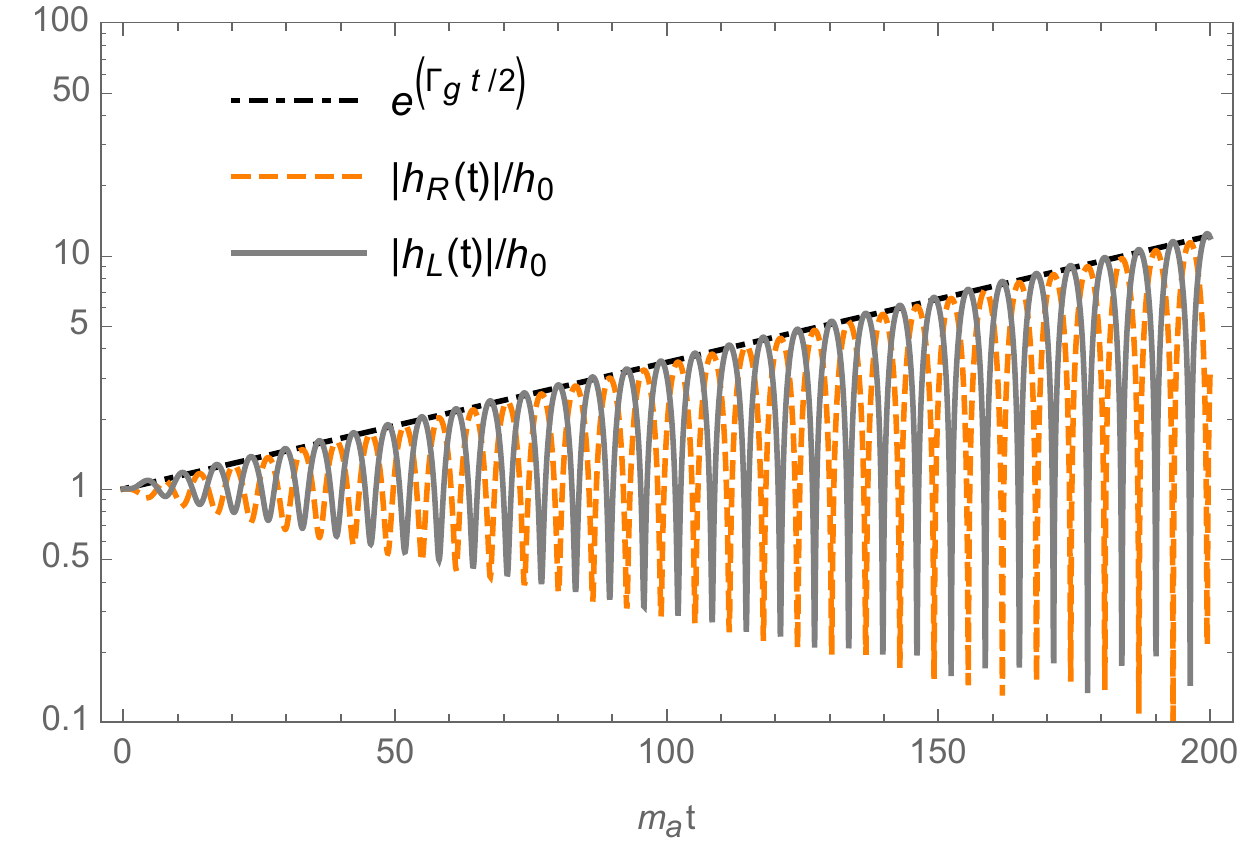} \\%\qquad\qquad
\includegraphics[scale=0.5]{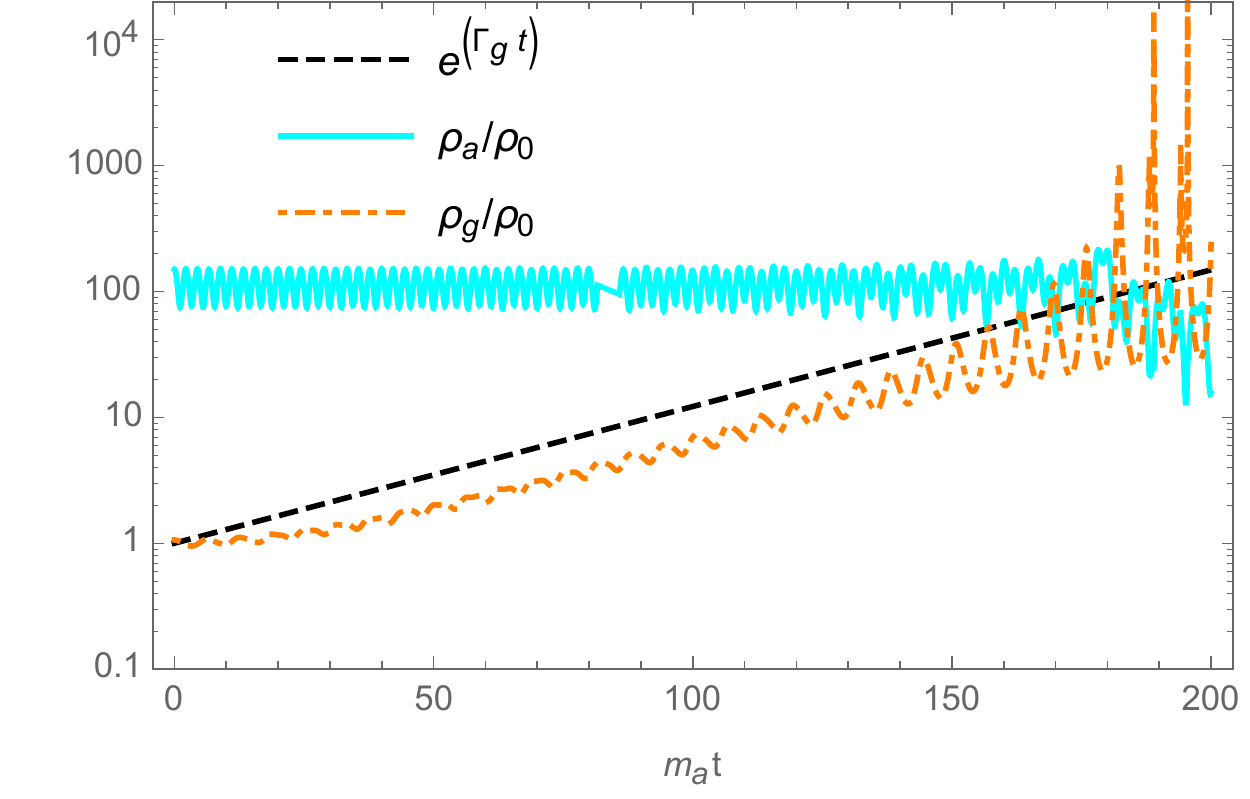}
\caption{Schematic diagrams for the amplification of gravitational waves, where $\Gamma_{\hg}/{m_{\va}}= 1/40$ in \eqref{GammaG} has been chosen in the numerical study. 
Top: the amplitude growth of the strength of the gravitational wave, which is normalized by the initial value of $h_0$. 
Bottom: the growth of energy densities of axion $\rho_{\va}$ and GW $\rho_{\hg}$, which are normalized by the initial value of GW $\rho_0\equiv \rho_g(t=0)$.  Notice the different growth phase for the $R, L$ modes of the gravitational wave is due to the parity breaking of the gravitational CS term. } \label{fig2}
\end{figure}

Now we can discuss the bursts of the gravitational waves around the axion dense object or the axion clouds around the black hole. We use the conventions in \cite{Yoshida:2017cjl,Yoshida:2017ehj}, and the typical peak frequency of such GW is around ${{\ka}} ={m_{\va}}/2$ with a bandwidth $\kappa_4 {\lr}{\vt_0} \frac{ m_{\va}^3  }{4}$. 
It corresponds to the process $\va \rightarrow{\hg\hg}$, where an axion decays into two gravitons. This is a different and more drastic process comparing to two axion annihilation to the graviton process proposed in \cite{Arvanitaki:2012cn,Ito:2019wcb}, where no gravitational {CS} coupling is assumed. 

The duration of the energy released in such burst is related to the axion objects and clouds properties. In the {FRB} cases, time duration is very short in milliseconds and the energy released is around $10^{38} \sim 10^{40}$ ergs with high flux densities \cite{Tkachev:2014dpa,Eby:2017azn,Fairbairn:2017sil,Helfer:2016ljl,Vaquero:2018tib}. If we assume the same clumps formed by axions also has the gravitational {CS} couplings, a portion of the axion energy, if not all, can be released to the gravitational waves. Notice here these two couplings are not related to each other, without assuming certain UV theory. If we assume such axion-like particles only have the gravitational couplings rather than photon couplings, we can expect the gravitational wave bursts with the similar total energy released as the {FRBs}, although the duration might be different, depending on the {CS} coupling.

\subsection{The amplification factor}
The amplification factor is another important parameter of such phenomenon, especially for the expected signal to background ratios. The gravitational waves traveled through the axion objects pick up an exponential factor, while the other gravitational waves remain the background. We can estimate that for the peak frequency, the exponentiated factor is $e^{\Gamma_{\hg} t_{\hg}}$, where 
 \begin{align}\label{GammaG}
 \Gamma_{\hg}=\kappa_4 {\lr}{\vt_0} \frac{ m_{\va}^3  }{4},\qquad t_{\hg} \simeq\frac{1}{{m_{\va}} {v_{\va}}c}.
\end{align}
For an estimation of the order, we have
 \begin{align}\label{Gammah}
\frac{\Gamma_{\hg}}{m_{\va}}&\sim \Big(\frac{{{\kappa_4}{\lr}}}{1\text{eV}^{-3}}\Big)\Big( \frac{m_{\va}}{10^{-9}\text{eV} } \Big)^2\Big( \frac{{\vt_0}}{10^{9}\text{GeV}} \Big).
\end{align}

We can now see that the observation of such bursts can put severe constraints on the gravitational {CS}  coupling and axion properties. We take the current upper bound of {CS} modified gravity coupling from \cite{Alexander:2009tp},
that $\ell \simeq 10^{11} \text{m}\simeq 5\times 10^{17}  { \text{eV}^{-1}}$.
In our notations, it turns out to be 
\begin{align}\label{lambdaG}
{\kappa_4}{\lr} &=\frac{\ell^2}{M_P}< 10^8 \text{eV}^{-3},\\
 {\lr} &=\frac{\ell^2 M_P}{2} < 10^{38} \text{eV}^{-1}.
\end{align}
Thus, if taking ${\kappa_4}{\lr}\sim 10^{-4}\text{eV}^{-3}$, ${m_{\va}}\sim \mu$eV and ${\vt_0}\sim 10^{9}\text{GeV}$ in \eqref{Gammah}, we can see that the current upper bound of $\frac{\Gamma_{\hg}}{m_{\va}}$ is around $10^2$, which means a very large amplification factor $e^{\Gamma_{\hg} t_{\hg}}\sim e^{100/{v_{\va}}}$, considering that ${v_{\va}}\sim 10^{-3}$. In Figure \ref{fig2}, we plot the schematic diagrams of the amplification of the gravitational wave. The energy density of the photons $\rho_{\hg}=T_{tt}^{(\hg)}$ are read out from the stress energy tensor of the gravitational wave $T_{\mu\nu}^{(\hg)}=\frac{c^4}{4\kappa_4}(\partial_\mu {h_{ij}})(\partial_\nu{h^{ij}}) $.

\subsection{Various Gravitational waves signals from axions}
%\label{Secother}
%Discussions on v
%\vspace{5pt}
%{\it\bf Various Gravitational waves signals from axions.} ---

For the early universe, one thing we would like to comment here is the gravitational wave production from $\va F\tilde{F}$ or $\va R\tilde{R}$ terms during inflation. They are at the frequency of $10^{-16}$ Hz to $10^{-10}$ Hz, which has been discussed in \cite{Alexander:2009tp} and \cite{Barnaby:2010vf,Barnaby:2012xt,Pajer:2013fsa,Mukohyama:2014gba} for the interests of the primordial gravitational waves. For both cases, one can solve the EM or GW modes in an FRW geometry and take the inflationary background. 
For the electromagnetic case with $\va F\tF$, the growing modes of the EM waves act like the sources for the background cosmological scalar fluctuations, which give rise to the near scale-invariant power spectrum and can be probed by the CMB and large scale structure. Especially, the tensor to scalar ratio becomes
$r_{\xi}=8.1\times10^7 \frac{H^2}{M_p^2}\Big[1+4.3\times 10^{-7}\frac{H^2}{M_p^2}\frac{e^{4\pi \xi}}{\xi^6}\Big]$, where $\xi= \frac{{{\lf}}\dot{\va}}{2H}$.
Moreover, this tensor fluctuation can be detected in much smaller scales when these modes exit the horizon close to the end of inflation. And a stochastic gravitational background can be possibly detected by advanced LIGO/VIRGO~\cite{Crowder:2012ik, Barnaby:2011qe}. 
For the $\va R\tilde{R}$ term one has the gravitational wave production directly, one needs to solve the field equations in an inflationary background. In the regime of $\Theta\equiv \frac{2{\lr} H^2}{M_p} \sqrt{2\epsilon}<10^{-5}$ with  the slow roll parameter  $\epsilon$, 
one can reach the tensor to scalar ratio
$\frac{r_\text{CS}}{r_\text{non-CS}} \sim1+0.022 \Theta^2$.
For the larger $\Theta$ the analysis has some technical difficulties, for example, the ghosts arise and the result is unknown currently~\cite{Alexander:2009tp, Alexander:2004us}. % from solving \eqref{GWFE}

%\vspace{-10pt}
\section*{ Acknowledgments}
%\vspace{-10pt}
{\small %\footnotesize
This work is support by National Natural Science Foundation of China (NSFC) under Grant No.12005255,  the Key Research Program of the Chinese Academy of Sciences (Grant No. XDPB15).  
S. Sun was supported by MIUR in Italy under Contract(No. PRIN 2015P5SBHT) and ERC Ideas Advanced Grant (No. 267985) \textquotedblleft DaMeSyFla";  Y. -L. Zhang was supported by the Grant-in-Aid for JSPS international research fellow (18F18315).
We thank a lot to T. Noumi and G. Shiu for the collaboration at the initial stage, as well as R. -G. Cai, S. Mukohyama, M. Peloso, J. Soda, B. Zhang for many helpful conversations. We also benefit from the talks at the ICTP workshop on ``Challenges and Opportunities of High Frequency Gravitational Wave Detection''(Oct.14-16, 2019  \href{http://indico.ictp.it/event/9006/}{http://indico.ictp.it/event/9006}).  
}

\end{document}